\documentclass{article}
\makeatletter

\@addtoreset{equation}{section}
\makeatother

\newtheorem{theo}{Theorem}[section]

\newtheorem{propo}[theo]{Proposition}

\newtheorem{defi}[theo]{Definition}

\newcommand{\be}{\begin{equation}}
\newcommand{\ee}{\end{equation}}

\font\ddpp=msbm10 
\def\R{\hbox{\ddpp R}}    

\def\C{\hbox{\ddpp C}}

\newcommand{\ben}{\begin{enumerate}}
\newcommand{\een}{\end{enumerate}}
\newcommand{\bit}{\begin{itemize}}
\newcommand{\eit}{\end{itemize}}

\title{{\bf On the Geometry of PP-Wave Type Spacetimes}}

\author{J.L. FLORES$^a$,
M. S\'ANCHEZ$^b$ \\
\\
{\small jflores@ugr.es, sanchezm@ugr.es}\\
{\small $^a$Department of Mathematics}, \\ 
{\small Stony Brook University,}\\
{\small Stony Brook, NY 11794-3651, USA,}\\
{\small $^b$Departamento de Geometr\'{\i}a y Topolog\'{\i}a}\\
{\small Facultad de Ciencias, Universidad de Granada}\\
{\small Avenida Fuentenueva s/n,
18071 Granada, Spain}\\
}

\date{}

\begin{document}

\textwidth=140mm
\textheight=200mm
\parindent=5mm

\maketitle

\begin{center} {\small \bf Abstract.} \end{center}

%

{\small

Global geometric properties of product manifolds ${\cal M}= M \times \R^2$, endowed with a metric type $\langle \cdot , \cdot \rangle = \langle \cdot , \cdot \rangle_R + 2 dudv + H(x,u) du^2$ (where
$\langle \cdot , \cdot \rangle_R$ is a Riemannian metric on $M$ and $H:M \times \R \rightarrow \R$ a function), which generalize classical plane waves, are revisited.  Our study covers causality (causal ladder, inexistence of horizons), geodesic completeness, geodesic connectedness and existence of conjugate points. Appropiate mathematical tools for each problem are emphasized and the necessity to improve several Riemannian (positive definite) results is claimed.

The behaviour of $H(x,u)$ for large spatial component $x$ becomes
essential, being a spatial quadratic behaviour  critical for many
geometrical properties. In particular, when $M$ is complete, if
$-H(x,u)$ is spatially subquadratic,  the spacetime becomes
globally hyperbolic and geodesically connected. But if a quadratic
behaviour is allowed (as happen in plane waves) then both global
hyperbolicity and geodesic connectedness maybe lost.

From the viewpoint of classical General Relativity, the properties which
remain true under generic hypotheses on ${\cal M}$ (as
subquadraticity for $H$) become meaningful. Natural assumptions on
the wave -finiteness or asymptotic flatness of the front- imply
the spatial subquadratic behaviour of $|H(x,u)|$ and, thus, strong
results for the geometry of the wave. These results not always
hold for plane waves, which appear as an idealized non-generic
limit case.}

\section{Introduction}

Among the reasons which contribute to the recent interest on pp-wave type spacetimes, we remark\footnote{Of course, there is also another influential reason: the possibilities of direct detection of gravitational waves. Hulse and Taylor were awarded with the Nobel prize in 1993 for the discovery in the seventies of  undirect evidences of their existence -a binary system loses an exact amount of rotational energy which can be conceived only as originated by gravitational waves. Nowadays, experimentalists look for direct evidences, and a generation of large scale interferometers is close to be operative throughout all the world (VIRGO, LIGO, GEO300, TAMA300...) and even the space (LISA). Although experimentalists' problems are very different to the ones in this paper, if they succeed, an excellent stimulous on waves for the whole relativistic community (and even for the curiosity of general public) will be achieved.}, on one hand, classical geometrical properties and, on the other, applications to string theory. About the former, pp-waves spacetimes, and specially plane waves, \cite{Br, ER, MTW} have curious and intriguing properties, which yielded questions still open or only recently solved. The well-known Penrose limit \cite{Pen-lim} (see also \cite{BlauPL,BlauPLSuper}) associates to every spacetime and choice of (unparametrized) lightlike geodesic a plane wave metric. Penrose \cite{Pen-pp} also emphasized that, in spite of being geodesically complete, plane waves are not globally hyperbolic (see Section \ref{s2} for definitions). Ehlers and Kundt \cite{EK} 
conjectured that gravitational plane waves are the only complete gravitational pp-waves. As we will see, now the lack of global hyperbolicity can be well understood, but Ehlers-Kundt conjecture still remains open. The applications to string theory have highlights as: (a) gravitational pp-waves are relevant spacetimes with vanishing scalar invariants (VSI, see \cite{Prav-pp, Prav-vsi} for a classification), and such spacetimes yield exact backgrounds for string theory (vanishing of $\alpha'$ corrections, \cite{HR-1,HR-2}), (b) Berenstein, Maldacena and Nastase \cite{HR-BMN} have recently proposed and influential solvable model for string theory by taking the Penrose limit in AdS$_5\times S^5$ spacetimes, or (c) after realizing that G\"odel like universes can be supersimetrically embedded in string theory, it was realized and emphasized that  these solutions were $T$-dual to compactified plane wave backgrounds \cite{MS-1,MS-2,MS}.

The necessity to understand better the geometry of waves and their potential applications to string theory, justify to study pp-waves from a wider perspective, where new mathematical tools appear naturally. The authors, in collaboration with A.M. Candela \cite{CFSgrg}, considered the following class of spacetimes, say PFW, which essentially include classical pp-waves (and, thus, plane waves):

\begin{equation}\label{pfw}
\begin{array}{c}
({\cal M},\langle\cdot,\cdot\rangle) \quad
 {\cal M} = M \times \R^2 \\
\langle\cdot,\cdot\rangle = \langle\cdot,\cdot\rangle_R
+ 2\ du\ dv + H(x,u)\ du^2 ,
\end{array}
\ee
where $(M,\langle\cdot,\cdot\rangle_R)$ is any smooth Riemannian ($C^\infty$, positive-definite, connected) $n$-manifold,  the variables $(v,u)$ are the natural coordinates
of $\R^2$ and the smooth scalar field
$H : M \times \R \to \R$ is not identically 0.

Our initial motivation to study such metrics came from some works
by two contributors to this meeting, R. Penrose and P.E. Ehrlich.
Penrose \cite{Pen-pp} showed that, even though plane waves are
strongly causal, they are not globally hyperbolic. Moreover, they
present a property of focusing of lightlike geodesics which
forbides not only global hyperbolicity but also the possibility to
embed them isometrically in higher dimensional semi-Euclidean
spaces. This is a {\em remarkable} property of plane waves but, as
he pointed out, it is also interesting to know ``whether the
somewhat strange properties of plane waves encountered here will
be present for waves which approximate plane waves, but for which
the spacetime is asymptotically flat, or asymptotically
cosmological in some sense''. Under our viewpoint, this is a
relevant question because the  geometrical properties of an exact
solution to Einstein's equation (as plane waves) are physically
meaningful only if they are ``stable'' in some sense -surely, not
fulfilled by a term as $H$ in (\ref{rr}). Even more, in the
setting of Penrose's  Strong Cosmic Censhorship Hypothesis \cite{Pe-scch}, generic
solutions to Einstein's equation with reasonable matter and
behaviour at infinity must be globally hyperbolic. And, obviously,
plane waves fail to be generic and well behaved at infinity
because of the many symmetries of the term $H$ (as well as the
part $M= (\R^2, dx^2 + dy^2)$).

Ehrlich and Emch, in a series of papers \cite{EE1, EE2, EE3} (see
also \cite[Ch. 13]{BEE}), carried out a detailed investigation of
the behaviour of all the geodesics emanating from a (suitably
chosen) point $p$ in a  gravitational plane wave. Then, they
showed that gravitational plane waves are causally continuous but
not causally simple, and characterized points necessarily
connectable by geodesics (see Subsection \ref{s5a}). Nevertheless,
again all the study relies on the ``non-generic nor stable''
conditions of symmetry of the gravitational wave, and the very
special form of $H(x,u)$: independence of the choice of the point
$p$,  explicit integrability of geodesic equations,  equal
equations for Killing fields, Jacobi fieds and geodesics...

In this framework, our goals in \cite{CFSgrg, FScqg}  were,
essentially: (i) to introduce the class of reasonably generic
waves (\ref{pfw}), (ii) to justify that, for a physically
reasonable asymptotic behaviour of the wave, $|H(\cdot , u)|$ must
be ``subquadratic'' (plane waves lie in the limit quadratic case),
and (iii) to show that, in this case, the geometry of the wave
presents good global  properties: global hyperbolicity
\cite{FScqg} and geodesic connectedness \cite{CFSgrg}. Even more,
the unstability of these geometric properties in the quadratic
case implied interesting  questions in Riemannian Geometry, studied in \cite{CFSjde}.

In the present article, we explain the role of the mathematical tools introduced in \cite{CFSgrg, CFSjde, FScqg}   in relation to both, classical problems on waves as \cite{Pen-pp, EK}, and posterior developments \cite{HRR-inher, HR-caus-pp, HR-hor-rev, HR-nohor, MS, Seno}. The proofs are referred to the original articles, or sketched in the case of further results.

This paper is organised as follows:

In Section \ref{s2}, some general properties of PFW's are
explained, including questions related to curvature and the energy
conditions. Remarkably, we  justify that the behaviour of
$|H(x,u)|$ for large $x$ must be subquadratic if the wave is
assumed to be finite or with fronts asymptotically flat in any
reasonable sense. This becomes relevant from the viewpoint of
classical General Relativity, and the global geometrical
properties of PFW's will depend dramatically on the possible
quadratic behaviour of $H$ or $-H$.

In Section \ref{s3}, we show that the behaviour of all the causal
curves can be essentially controlled in a PFW (the more accurate
control for existence of causal geodesics is postponed to Section \ref{s5}). In
Subsection \ref{s3a} a detailed study of the causal hierarchy of
PFW's is carried out. In particular, Theorem \ref{tgh} answers
above Penrose's question, by showing that the causal hierarchy of plane
waves is ``unstable'' or ``critical'': deviations in the
superquadratic direction of $-H$ may transform them in
non-distinguishing spacetimes, but deviations in the (more
realistic) subquadratic direction yield global hyperbolicity.
Posterior results by Hubeny, Rangamani and Ross \cite{HRR-inher}
are also discussed. In Subsection \ref{s3b} the criterion on
inexistence of horizons posed by Hubeny and Rangamani
\cite{HR-nohor, HR-hor-rev} is explained, and a simple proof
showing that it holds for any PFW is given.

In Section \ref{s4}, geodesic completeness is studied. We claim that this problem is equivalent to a purely Riemannian problem  (Theorem \ref{th3.2}), which has been solved satisfactorily only for autonomous $H$, i.e., $H(x,u) \equiv H(x)$. The power of the known autonomous results (which yield completeness for at most quadratic $H(x)$, Theorem \ref{tcom}) is illustrated by comparison with the examples in \cite{HR-caus-pp}. Then, we claim the necessity to improve the non-autonomous ones. Moreover, Ehlers-Kundt conjecture deserves a special discussion. Even though easily solvable under  at most quadraticity  for $x$ (Theorem \ref{tek}), it remains open in general.

In Section \ref{s5}, the problems related to geodesic connectedness are studied. The key is to reduce the problem to a purely Riemannian problem, in fact, the classical variational problem of finding critical points for a Lagrangian type kinetic energy minus (time-dependent) potential energy. That is, to solve this classical problem becomes equivalent to solve the geodesic connectedness problem in  PFW's. Remarkably, in order to obtain the optimal results on waves (extending Ehrlich-Emch's ones) we had to improve the known Riemannian results; in the Appendix, this Riemannian problem is explained. Finally, the existence of conjugate points is discussed, and reduced again to a purely Riemannian problem. Energy conditions tend to yield conjugate points for causal geodesics. But, in agreement with the remainder of the results of the present paper, the above mentioned focusing property of lighlike geodesics in plane waves becomes highly non-generic.

\section{General properties of the class of waves}\label{s2}

\subsection{Definitions}

Let us start with some simple properties of the metric
(\ref{pfw}). The assumed geometrical background  can be found in
well--known books as \cite{BEE,HE,ON} and, following \cite{ON}. Vector 0 will be regarded as spacelike instead of lightlike.

Vector field $\partial_v$ is  parallel and lightlike, and the
time-orientation will be chosen to make it past-directed. Thus,
for any future directed causal curve $z(s)=(x(s),v(s),u(s))$,
\[
\dot u(s) = \langle \dot z(s), \partial_v\rangle \geq 0,
\]
and the inequality is strict if $z(s)$ is timelike. As $\nabla u
=\partial_v$, coordinate $u: {\cal M} \rightarrow \R$ makes the
role of a ``quasi-time'' function \cite[Def. 13.4]{BEE}, i.e., its
gradient is everywhere causal and any causal segment $\gamma$ with
$u\circ \gamma$ constant (necessarily, a lightlike pregeodesic
without conjugate points except at most the extremes) is
injective. In particular, the spacetime is causal (see also Section
\ref{s3a}). The hypersurfaces $u\equiv$ constant are degenerate,
with radical Span$\partial_v$. The hypersurfaces
($n$-submanifolds) of these degenerate hypersurfaces which are
transverse to $\partial_v$, must be isometric to open subsets of
$M$. The fronts of the wave (\ref{pfw}) will be defined as the
(whole) submanifolds at constant $u, v$.

According to  Ehlers and Kundt \cite{EK} (see also \cite{Bi}) a vacuum
spacetime is a plane-fronted gravitational wave if it contains a shearfree
 geodesic lightlike vector field $V$, and  admits ``plane waves'' --spacelike (two-)surfaces  orthogonal to $V$.
The best known subclass of these waves are the (gravitational)
``plane-fronted waves with parallel rays'' or  pp-waves, which are
characterized by the condition that
 $V$ is covariantly constant $\nabla V=0$.
Ehlers and Kundt gave several  characterizations of these
waves in coordinates, and we can admit as definition of a pp-wave, the spacetime
 (\ref{pfw}) with $M= \R^2$. The pp-wave is  gra\-vi\-ta\-tio\-nal (i.e., vacuum, see Subsection \ref{s2b}) if and only if the ``spatial'' Laplacian $\Delta_xH(x,u)$ vanishes. Plane waves constitute the (highly symmetric) subclass of pp-waves with $H$ exactly quadratic in $x$ for appropiate global coordinates on each front; that is, when we can assume:
\be \label{rr}
H(x,u) = (x^1,x^2) \left(
\begin{array}{rr}
f_1(u) & g(u) \\
g(u) & -f_2(u)
\end{array}
\right)
\left(
\begin{array}{l}
x^1 \\
x^2
\end{array}\right)
\ee where $f_1, f_2, g$ are arbitrary (smooth) functions. When
$f_1\equiv f_2$, the plane wave is gravitational, and there are
other well-known subclasses (sandwich plane wave if $f_1, f_2, g$
have compact support; purely electromagnetic plane wave if
$f_1\equiv -f_2$, $g\equiv 0$,  etc.)

Recall that, in our type of metrics (\ref{pfw}), no restriction on
the Riemannian part $(M,\langle\cdot,\cdot\rangle_{R})$ is
imposed. This seems convenient for different reasons as, for
example: (i) the generality in the dimension $n$,  for
applications to strings, (ii) the generality in the topology, for
discussions on horizons, or (iii)  the generality in the metric,
to obtain ``generic results'', not crucially dependent on very
special particular properties of the metric. In this ambient, a
name as generalized pp-wave or ``{\em $M$-fronted wave with
parallel rays}'' ({\em $M$p-wave})  seems natural for our
spacetimes (\ref{pfw}). Nevertheless, we will maintain the name
PFW (plane fronted wave) in agreement with previous references
\cite{CFSgrg, FScqg} or the nomenclature in \cite{BEE}, but with no
further pretension.

\subsection{Curvature and matter} \label{s2b}

Fixing some local coordinates $x^1, \dots, x^n$ for the Riemannian
part $M$, it is straightforward to compute the Christoffel symbols
of $\langle\cdot,\cdot\rangle$ and, thus, to relate the
Levi-Civita connections $\nabla, \nabla^R$ for ${\cal M}$ and $M$
resp. (see \cite{CFSgrg}). We remark the following facts:

\bit \item $M$ is totally geodesic, i.e., $\nabla_{\partial i}
\partial_j = \nabla^R_{\partial_{i}} \partial_j $, $i,j=1, \dots, n$. \item
$2\nabla_{\partial_u } \partial_u = - $grad$_xH[x,u] +
\partial_uH(x,u) \partial_v ; \quad$ $2\nabla_{\partial i}
\partial_u =  \partial_iH(x,u) \partial_v$; thus, the curvature
tensor satisfies \be \label{hessH} -2 R(\cdot, \partial_u ,
\partial_u, \cdot ) = \mbox{Hess}_xH[x,u](\cdot, \cdot). \ee Here
grad$_xH$ ($\equiv \nabla_xH$, in what follows) and Hess$_xH$
denote the spatial (or ``trasverse'') gradient and Hessian of $H$, respectively.
\item The Ricci tensors of ${\cal M}$ and $M$ satisfy
\[
{\rm Ric} = \sum_{i,j=1}^n R^{(R)}_{i j} d x^i \otimes d x^j -\frac{1}{2}\Delta_{x}H du \otimes du .
\]
 Thus,  Ric is null if and only if both, the Riemannian
Ricci tensor Ric$^{(R)}$ and the spatial Laplacian $\Delta_{x}H$, vanish.
\eit
From the last item, it is easy to check that the timelike condition condition holds if and only if
$$ \mbox{Ric}^{(R)}(\xi, \xi)\geq 0, \quad \Delta_{x}H \leq 0 ,\quad\hbox{for all}\;\; x\in M,\;\;\xi\in T_{x}M.$$
Even more, in dimension 4 all the energy conditions are equivalent
and easily characterized \cite[Proposition 5.1]{FScqg}:
\begin{propo} \label{45a}
Let ${\cal M} =  M \times \R^2$ be a 4-dimensional PFW, and let
$K(x)$ be the (Gauss) curvature of the 2-manifold $M$. The
following conditions are equivalent:

(A) The strong energy condition (Ric$(\xi, \xi)\geq 0$ for all timelike $\xi$).

(B) The weak energy condition ($T(\xi, \xi)\geq 0$ for all timelike $\xi$).

(C) The dominant energy condition ($-T_b^a \xi^b $ is either 0 or
causal and future-pointing, for all future-pointing timelike $\xi\equiv \xi^b$).

(D) Both inequalities:
$$K(x) \geq 0 , \quad \Delta_xH(x,u) \leq 0, \quad \quad  \forall (x,u) \in M\times \R.$$

\end{propo}

\subsection{Finiteness of the wave and decay of $H$ at infinity}

Now, let us discuss {\em minimum necessary} assumptions which must be satisfied by a PFW, if it is supposed  ``finite'' or ``asymptotically vanishing'' in any reasonable sense. In principle, one could think that $M$ should be asymptotically flat, but we will not impose this strong condition a priori (say, non-trivial  fronts at ``cosmological scale'' are admitted). At any case, it would not be too relevant for our problem: plane waves have flat fronts, and are not by any means finite.

As we have said, all the scalar curvature invariants of a gravitational pp-wave vanish. Thus, instead of such scalars, we will focus on the spatial Hessian Hess$_xH$. In the case of plane waves, Hess$_xH$ is essentially the matrix in (\ref{rr}) -{\em transverse frequency matrix} of the lightlike geodesic deviation \cite{Pen-pp}.
By equality (\ref{hessH}), Hess$_xH$ is related to the most ``characteristic'' curvatures of the wave;  these curvatures -taken along a lightlike geodesic- admit an intrinsic interpretation in terms of Penrose limit (see \cite{BlauPL}, specially the discussions around formulas (1.2), (2.13)). According to \cite{HR-nohor, HR-hor-rev}, ``to go arbitrarily far'' in a pp-wave can be thought as taking $v, x$ large
for each fixed $u$  (see also the Subsection \ref{s3b}). Therefore, any sensible definition of finiteness or asymptotic vanishing of the PFW seems to imply that Hess$_xH[x,u]$ must go (fast) to 0 for large $x$.

Rigourously, let $\lambda_i(x,u), i=1, \dots ,n$, be the eigenvalues of Hess$_xH[x, u]$,
$d(\cdot, \cdot)$ the Riemannian distance on $M$ and fix any $\bar x \in M$. From the above discussion, if the wave vanishes asymptotically  then
$\lim_{d(x, \bar x)\rightarrow \infty} \lambda_i(x,u)$ must vanish fast for each $u$. Therefore, putting $|\lambda(x,u)|$ equal to the maximum of the $|\lambda_i(x,u)|$'s, we can assume as definition of asymptotic vanishing for a PFW:
\be \label{lambda} 
|\lambda(x,u)| \leq \frac{A(u)}{d(x,\bar x)^{q(u)}} 
\ee
for some continuous functions, $A(u)$ and $q(u)>0$.

Inequality (\ref{lambda}) implies bounds for the spatial growth of  $|H|$, as the next proposition shows. But, first, let us introduce the following definition. Let $V(x,u)$ be a continuous function $V: M\times \R \rightarrow \R$. We will say that $V(x,u)$ behaves {\em subquadratically at spatial infinity} if
$$V(x,u) \le R_1(u) d^{p(u)}(x,\bar x) + R_2(u) \quad
\forall (x,u) \in M \times \R,
$$
for some continuous functions $R_1(u), R_2(u) (\geq 0)$ and $p(u) < 2$. If the last inequality is relaxed in $p(u) \leq 2, \forall u \in \R$ then $V(x,u)$ behaves {\em at most quadratically at spatial infinity}. Now, we can assert the following result (see \cite[Proposition 5.3]{FScqg} for the idea of the proof -notice that the completeness of $M$ is not necessary now, as any curve can be approximated by broken geodesics).

\begin{propo}  If the PFW vanishes asymptotically as in (\ref{lambda}), then $|H(x,u)|$ behaves subquadratically at spatial infinity.
\end{propo}
It must be emphasized:
\begin{enumerate}
\item The asymptotic vanishing condition (\ref{lambda}) implies
subquadraticity for $|H(x,u)|$, but the converse is not true.  In
the remainder of this paper, we will use only this more general
subquadratic condition or, even, only the subquadraticity (or at most
quadraticity) of $H$ or $-H$. So, the range of application of our
results will be wider. \item Of course, inequality (\ref{lambda})
is compatible with the energy conditions. A simple explicit
example can be constructed by taking   $H(x^1,x^2,u) \equiv H_u(x), (x\equiv x^1)$ with:
$$ -\frac{A(u)}{|x|^{q(u)}}  \leq \frac{d^2H_u}{dx^2}(x)  \leq 0$$
for some $A(u), q(u) >0$.
\item For  plane waves,  neither $H$ nor $-H$ behaves subquadratically. In fact, the eigenvalues of Hess$_xH[x,u]$ are independent of $x$, and the fronts of the wave are not ``finite''. This is a consequece of the idealized symmetries of the front waves. Nevertheless, $|H(x,u)|$ behaves at most quadratically at infinite and, thus, plane waves lie in the limit quadratic situation.
\end{enumerate}

\section{Causality} \label{s3}
\subsection{Positions in the causal ladder}  \label{s3a}

Recall first the causal hierarchy of spacetimes \cite{BEE}:

\begin{center}
Globally hyperbolic  $\Rightarrow$ Causally simple $\Rightarrow$ Causally continuous \\ $\Rightarrow$ Stably causal
$\Rightarrow$ Strongly causal \\ $\Rightarrow$ Distinguishing $\Rightarrow$ Causal $\Rightarrow$ Chronological
\end{center}
Roughly, a spacetime is causal if it does not contain closed causal curves, strongly causal if there are no ``almost closed'' causal curves and stably causal if, after opening slightly the light cones, the spacetime remains causal. It is widely known that stable causality is equivalent to the existence of a {\em continuous} time function (see \cite{HE, BEE}), but only recently the existence of a {\em smooth} time function with nowhere lightlike gradient -i.e., a ``temporal'' function- has been proven \cite{BS2} (see also \cite{BS3} for the history of the problem). Globally hyperbolic spacetimes can be defined as the strongly causal ones with compact diamonds $J^+(p)\cap J^-(q)$ for any $p, q$. They were characterized by   Geroch as those possesing a Cauchy hypersurface (which can be also chosen smooth and spacelike \cite{BS1}). PFW's are always causal (Section \ref{s2}) and the following result was proven in \cite{FScqg}:

\begin{theo}\label{tgh}
Any PFW with $M$ complete and $-H$ spatially subquadratic is globally hyperbolic.
\end{theo}
The following points must be emphasized: \ben \item The proof is
carried out by showing strong causality and the compactness of the
diamonds. From the technique, one can also check that, if $-H$ is
at most quadratic at spatial infinity, then the spacetime is
strongly causal (with no assumption on the completeness of $M$).
\item Hubeny, Rangamani and Ross \cite{HRR-inher} constructed
explicitly a temporal function for  plane waves. As the light
cones of an at most quadratic pp-wave can be bounded by the cones
of a plane wave, they claim that {\em any pp-wave with $-H$
 at most quadratic at spatial infinite is stably causal}. (They also use the
temporal function to study  quotients of the wave by the isommetry
group generated by a spacelike Killing field, which maybe stably
causal or non-chronological,  see also \cite{MS}).

Recall also from the Introduction, that  gravitational plane waves are causally continuous (the set valued maps $I^\pm$ are outer continuous) but not causally simple (because the causal future or past of a point may be non-closed).

\item If $-H(x,u)$ were not at most quadratic, then the spacetime may be even non-distinguishing (the chronological future or past of two distinct points are equal). In fact, a wide family of non-distinguishing examples with $-H$ ``arbitrarily close'' to at most quadratic  (and $M$ complete) is constructed in \cite[Proposition 2.1]{FScqg}; in these PFW's, the chronological futures  $I^+(x,v,u)$ depend only of $u$. In particular, any pp-wave such that  $-H$ behaves as $|x|^{2+\epsilon}, \epsilon>0$ for large $|x|$ is non-distinguishing \cite[Example 2.2]{FScqg}.

Nevertheless, of course, the spatially subquadratic or at most quadratic behaviours of $-H$ are not necessary for  global hyperbolicity or strong/ stable causality, as explicit counterexamples \cite[Example 4.5]{FScqg} show (compare with \cite[Section 4]{HRR-inher}).

\item A curious phenomenon suggested in \cite[Section
4]{HRR-inher}, is that the class of distinguishing but non-stably
causal pp-waves (or even PFW's) might be empty. In fact, our technique in
\cite{FScqg} showed that if the class were non-empty, then it would not be too significative.





\een The technique involved for Theorem \ref{tgh} can
be understood as follows. Any future-directed timelike curve
$\alpha$ can be reparametrized by the quasi-time $u$: $\alpha(u) =
(x(u), v(u), u), u \in [u_0, u_1].$ The proof is based on
inequalities which relate the distance covered by $x(u)$ with the
extreme points of $v(u)$. Say, fixed $\epsilon >0$, and
 $0<u_1-u_0\leq \epsilon$, $u \in [u_0, u_1]$, then:
$$
\frac{1}{\epsilon^{2}} \int_{u_{0}}^{u} d^{2}(x(s),x(u_{0})) ds
\leq \int_{u_{0}}^{u}\langle \dot{x}(s),\dot{x}(s)\rangle_R ds
$$
$$< 4\left(R_2'(u-u_0) - (v(u)-v(u_0))\right)
$$
where the constant $R_2' = R_2'(u_0, \epsilon)$ is independent of $x(u_0)$ in the subquadratic case (in the finer proof of strong causality for the at most quadratic case, $R_2'$ is allowed to depend on a compact subset where $x(u_0)$ lies, and $\epsilon >0$ is not fixed a priori). Then, such an inequality is used:

\bit \item For strong causality, to prove that, fixed a small
neighborhood of a point $z_0$ (which can be chosen ``square'' in the coordinates $u,v$), and any causal curve with extremes in this neighborhood, 
the restrictions on the extremes for $u, v(u)$ also imply
restrictions on the distance between $x(u), x(u_0)$. This forces
the whole $x(s)$ to remain in a small neighborhood. \item For global
hyperbolicity, to prove also that the projections of each diamond
$J^+(p) \cap J^-(q) \subset M \times \R^2$ on each factor $M,
\R^2$ are bounded for the natural (complete) Riemannian distances
$d$ on $M$ and $du^2+dv^2$ on $\R^2$. Therefore $J^+(p) \cap
J^-(q)$ will be included in a compact subset, which turns to yield
compactness. \eit

\subsection{Causal connectivity to infinity and horizons}\label{s3b}
Next, let us comment the applicability of these techniques to the study  of horizons in PFW's. The possible existence of horizons in gravitational pp-waves and, in general, in vanishing scalar curvature invariant spacetimes, have attracted interest recently. Hubeny and Rangamani \cite{HR-nohor, HR-hor-rev} proposed a criterion for the existence of horizons in pp-waves, and they proved the {\em inexistence } of such horizons. In a more standard framework, Senovilla \cite{Seno} proved the inexistence of closed trapped or nearly trapped surfaces (or submanifolds in any dimension) in VSI spacetimes. Next, we will give a simple proof of the inexistence of horizons, in the sense of Hubeny and Rangamani, for an arbitrary PFW.

Hubeny-Rangamani's criterion \cite[Sections 2.2, 4]{HR-nohor}  can
be reformulated as follows\footnote{There is a change of sign for
$v$ now respect to this  reference, because our convention for the
metric uses $dudv$ instead of $-dudv$}: {\em a pp-wave spacetime
(or, in general, any PFW) ${\cal M}$ does not admit an event
horizon if and only if, given any points $z_0= (x_0,v_0,u_0),
(x_1,v_1,u_1)\in {\cal M}$ with $u_0<u_1$, there is $-v_\infty >
-v_1$ such that a future-directed causal curve from $z_0$ to
$z_\infty= (x_1,v_\infty,u_1)$ exists}. According to the authors,
this criterion tries to formalize the intuitive idea that any
point of the spacetime is visible to an observer who is
``arbitrarily far''. In fact, one may think  $u_1$ as being close
to $u_0$, and $x_1$ as arbitrarily far from $x_0$.

To check that this criterion is satisfied for any PFW, choose any
curve $\alpha$ starting at $z_0$ parametrized by $u$, $\alpha(u) =
(x(u), v(u), u), u \in [u_0, u_1]$ such that $x(u_1)=x_1$. Putting
$E_\alpha (u) = \langle \dot \alpha (u), \dot \alpha (u)\rangle =
\langle \dot x (u), \dot x (u)\rangle_{R} +2\dot v (u)+
H(x(u),u)$, then function $v(u)$ can be reobtained from
$E_\alpha(u)$ as:
\[
v(u)-v_0=\frac{1}{2}\int_{u_{0}}^{u}\left( E_{\alpha}(\bar
u)-\langle \dot{x}(\bar u),\dot{x}(\bar u)\rangle_{R}- H(x(\bar
u),\bar u)\right) d\bar u, \quad \forall u\in [u_{0},u_1].
\]
Choosing $E_\alpha <0$ the curve $\alpha$ becomes timelike and future directed, and, as $|E_\alpha|$ can be chosen arbitrarily big (and even constant, if preferred), the value of $-v(u_1)$ can be taken arbitrarily big, as required.

\section{Geodesic completeness}\label{s4}

\subsection{Generic results} \label{s4a}
From the direct computation of Christoffel symbols, it is straightforward to write geodesic equations in local coordinates. Remarkably, the three geodesic equations for a curve
$z(s)= (x(s), v(s), u(s))$
can be solved in the following sequence \cite[Proposition 3.1]{CFSgrg}:
\begin{enumerate}
\item[$(a)$] $u(s)$ is affine, $u(s) = u_0 + s \Delta u$, for some $\Delta
u\in \R$.

\item[$(b)$] Then $x = x(s)$ is a solution on $M$ of
\[
D_s\dot x = - \nabla_x V_{\Delta}(x(s),s) \quad \mbox{for all $s
\in \ ]a,b[$,}
\]
where $D_s$ denotes the covariant derivative and
\[
V_{\Delta}(x,s) = -\ \frac{(\Delta u)^2}{2}\ H(x, u_0 + s \Delta u);
\]

\item[$(c)$] Finally, $v(s)$ can be computed from:
\[
v(s) = v_0 + \frac{1}{2 \Delta u} \int_0^s \left( E_z - \langle
\dot x(\sigma), \dot x(\sigma)\rangle_{R} + 2
V_{\Delta}(x(\sigma), \sigma)\right) d\sigma.
\]
where $E_z=\langle \dot z(s), \dot z(s)\rangle $ is a constant (if $\Delta u = 0$ then $v = v(s)$ is also affine).
\end{enumerate}
In particular, geodesic completeness is reduced, essentially, to the completeness of trajectories for (non-autonomous) potentials on $M$, and one can prove
\cite[Theorem 3.2]{CFSgrg}:

\begin{theo}
\label{th3.2}
A PFW is geodesically complete  if and only if the Riemannian manifold $M$
 is  complete  and the trajectories of
\[
D_s\dot x = \ \frac 1 2\ \nabla_x H(x,s)
\]
are also complete.
\end{theo}
Recall that the completeness of $M$ is an obvious necessary
condition (the wave fronts are totally geodesic) and, then, the
question is fully reduced to a purely Riemanian problem: the
completeness of the trajectories of the potential $V=-H/2$. This
problem was studied by several authors in the 70's \cite{Eb,Go,WM}
and they obtained very accurate results when the potential is
autonomous, i.e., $H$ independent of $u$. For example, a result by
Weinstein and Marsden \cite{WM} (see also \cite[Theorem
3.7.15]{AM} or \cite[Section 3]{CFSgrg}), formulated  in terms of
positively complete functions, yields as a straightforward
consequence:
\begin{theo} \label{tcom}
Any PFW with $M$ complete and $H(x,u) \equiv H(x)$ at most quadratic is geodesically complete.
\end{theo}
Recall that here only $H$ (and no $-H$) needs to be controlled. As
an example of the power of this result, one can check that the
explicit examples of pp-waves exhibited in \cite{HR-caus-pp},
which were proven to be complete (by integrating -decoupled-
geodesic equations), lie under the hypotheses\footnote{For the
comparison of hypotheses, recall that their function $F(x,u)$
plays the role of our $-H(x,u)$.} of Theorem \ref{tcom}. For
example, for PP1 (see \cite[Section 5.2]{HR-caus-pp}), $H(x,y)=
\cos y - \cosh x$; for PP2, $H= -\sum_j f_j(x^j)$ with the $f_j$'s
bounded from below; in both cases, $H$ is upper bounded. Even
more, their incomplete examples violate strongly the conditions of
Theorem \ref{tcom}. For example, for the monopole pp-waves in PP3
the Riemannian part $M$ may be incomplete, and for the example PP4
one has the highly violating coefficient $H(x,y)= -e^{y}\sin x$.

Nevertheless, the results for non-autonomous potentials  are not so accurate \cite{Go}. But this is the case of plane waves, which are geodesically complete in any dimension (see \cite[Proposition 3.5]{CFSgrg}) and, then,  to find general and accurate criteria seems an interesting topic to research.

\subsection{Ehlers-Kundt question}
From a fundamental viewpoint, the following  question on pp-waves ($M=\R^2$) was posed by Ehlers and Kundt \cite{EK} (see also \cite{Bi} or \cite{HR-caus-pp}):
\begin{quote}
Is any {\em complete} gravitational pp-wave a plane wave?
\end{quote}
As they pointed out, complete gravitational pp-waves represent graviton fields generated independently of matter (vacuum) or external sources (completeness). Then, they are the analogous to source-free photons in electrodynamics.

Notice that the hypotheses become relevant for both, the physical interpretation and the involved mathematical problem.  In fact, from Theorem \ref{th3.2} (with $V=-H/2$) and the fact that linear terms in the expression of $H$ in (\ref{rr}) can be dropped by choosing appropiate coordinates, previous question is equivalent to:
\begin{quote}
Let $V((x,y),s)$, $V: \R^2\times \R \rightarrow \R$ be an harmonic function in $(x,y)$. If the trajectories for $V$ as a (non-autonomous) potential on $\R^2$ are complete, must $V$ be a (harmonic) polinomyal of degree $\leq 2$ for each fixed $s$ (i.e., $V((x,y),s)= f(s)(x^2-y^2) + 2g(s)xy + c(s)x + d(s) y + e(s)$)?
\end{quote}
Notice that the harmonicity of $H$ allows to use techniques of
complex variable. In fact, it is  easy to show:

\begin{theo}\label{tek}
Any gravitational pp-wave such that $H(x,u)$ behaves  at most quadratically at spatial infinity is a (necessarily complete) plane wave.
\end{theo}
To prove it, put $\zeta= x+iy$, $H\equiv H(\zeta, u)$ and consider the complex function $f(\zeta, u)$ which is holomorphic in $\zeta$ with real part equal to $H$. Then, $f(\zeta, u)/\zeta^2$ is meromorphic for $\zeta \in \C$ and bounded for big $\zeta$. Thus, for each $u$, whenever $f(\zeta,u)$ is not constant, it presents a pole  at infinity of order $\leq 2$. That is,  $f(\cdot, u)$ is a complex polinomyal of degree at most $ 2$, and the result follows directly.

Even though Theorem \ref{tek} covers the most meaningful cases from the physical viewpoint (and is free of hypotheses on completeness), the above questions remain open as a mathematical problem with roots in the foundations of the theory of gravitational waves.

\section{Geodesic connectedness and conjugate points}\label{s5}

\subsection{The Lorentzian problem} \label{s5a}
Next, we will study geodesic connectedness of PFW's, that is, we
will wonder:  fixed any $z_0=(x_0, v_0, u_0), z_1=(x_1, v_1, u_1)
\in {\cal M}$,  is there any geodesic  connecting  $z_0, z_1$?
This problem becomes relevant from different viewpoints (see
\cite{Sa-cata} for a survey): (a) the connectivity of a point
$z_0$ with any point $z_1 \in I^+(z_0)$ through a timelike
geodesic, admits an obvious physical interpretation, and is
satisfied by all globally hyperbolic spacetimes (Avez-Seifert
result), (b) the geodesic connectedness of a Lorentzian manifold
-through geodesics of any causal type- is a desirable geometrical
property\footnote{Trivially satisfied for complete Riemannian
manifolds but not necessarily for complete Lorentzian ones, as de
Sitter spacetime.}, which admits a natural variational
interpretation and, then, yields an excellent motivation to study
critical points of indefinite functionals from a mathematical
viewpoint \cite{Ma}, (c) the possible multiplicity of connecting
geodesics is related to the existence of conjugate points.

These questions were studied by Penrose \cite{Pen-pp} and Ehrlich and Emch \cite{EE1, EE2, EE3} for plane waves, by integrating geodesic equations. They proved that there exists a natural concept of  {\em conjugacy for pairs} $u_0, u_1 \in \R$, $u_0<u_1$, and  obtained the following results:

\begin{enumerate}
\item (Penrose). Lightlike geodesics are focused when $u_0, u_1$ are conjugate (at least for ``weak'' sandwich waves). In this case, all the lightlike geodesics starting at $z_0$ (except one),
\begin{itemize}
\item either cross a fixed point with $u=u_1$ (anastigmatic
conjugacy, in electromagnetic plane waves) \item or cross a fixed
line (astigmatic conjugacy, in gravitational or mixed plane
waves).
\end{itemize}
\item (Ehrlich-Emch). The connectable points for astigmatic gravitational plane waves can be characterized in an accurate way:
\begin{itemize}
\item if $u_1$ lies before the first conjugate point of $u_0$, then there exists  an unique geodesic between $z_0$ and $z_1$, which is causal if $z_0 < z_1$.
\item otherwise, connecting geodesics may not exist and, in fact,  {\em gravitational plane waves are not geodesically connected}.
\end{itemize}
\end{enumerate}

\subsection{Relation with a purely Riemannian variational problem}

From the study of geodesic equations in Section \ref{s3}, and the classical relation between connecting trajectories for a potential and extremal of Lagrangians, it is not difficult to prove \cite{CFSgrg}:

\begin{propo} \label{prp} Fixed $z_0, z_1 \in {\cal M}$, they are equivalent:
\begin{enumerate}
\item[$(a)$] $z_0$ and $z_1$ can be connected by a geodesic.

\item[$(b)$] There exists a solution for the Riemannian problem
$$\left\{
\begin{array}{l}
D_s\dot x (s)= - \nabla_x V_{\Delta}(x(s),s)\quad \mbox{for all $s \in [0,1]$}\\
x(0) = x_0 ,\;\; x(1) = x_1, \\
\end{array}
\right. $$ where $V_{\Delta}(x,s) = -\ \frac{(\Delta u)^2}{2}\
H(x, u_0 + s \Delta u), \; \Delta u = u_1-u_0$.

\item[$(c)$] There exists a critical point for action functional
${\cal J}_{\Delta}$ defined on the space of absolutely continuous
curves $x:[0,1]\rightarrow M$ which connect $x_0, x_1$,
\be \label{jdelta}
{\cal J}_{\Delta}(x)= {1\over 2}\ \int_0^1 \langle\dot x,\dot x\rangle_{R}\ ds
\ - \int_0^1 V_{\Delta}(x,s)\ ds. \ee

\end{enumerate}
\end{propo}
Of course, $(c)$ is {\em the most classical problem in Lagrangian Mechanics}. Nevertheless (as a surprise for us), it had not been fully solved in the quadratic case. This case corresponds to plane waves and, thus, in order to obtain optimal Lorentzian results (reobtaining in particular Ehrlich-Emch's), we had to improve the known Riemannian  ones.
The final Riemannian result  \cite{CFSjde} is the following (see the Appendix for a discussion on the problem):

\begin{theo} \label{trp} Let $(M,\langle\cdot,\cdot\rangle_{R})$ be a
complete (connected) $n$--dimensional Riemannian manifold.
Assume that $V\in C^1(M \times [0,1],\R)$
is at most quadratic in $x$ in the following way:
$$V(x,s) \le \lambda d^2(x,\bar x)
+ \mu d^{p}(x,\bar x) + k
\quad \forall (x,s) \in M \times [0,1],$$
for some fixed point $\bar x \in M$ and constants $p <2$, $\lambda, \mu, k \geq 0$.

If $\lambda < \pi^2/2$ then, for all $x_0, x_1 \in m$, there
exists at least one critical point (in fact, an absolute minimum)
of ${\cal J}_{\Delta}$ in (\ref{jdelta}).
In particular, this happens if $V$ is subquadratic, i.e., when $\lambda =0$.

If, additionally, $M$
is not contractible, then there exists a sequence of critical points $\{x_k \}_k$
such that $$\lim_{k \to +\infty}{\cal J}_\Delta (x_k) \rightarrow +\infty.$$
\end{theo}
Notice that, in the  quadratic bound  of $V$, the smaller the constant $\lambda$, the stronger the conclusion. One can also assume that $\lambda$ (as well as  $p, \mu$) depend on $s$, and then take the maximum of $\lambda([0,1])$ for the conclusion.

\subsection{Optimal results for connectedness of PFW's}
Now, the application of Proposition \ref{prp} and Theorem \ref{trp} (plus a further discussion for the case of causal geodesics) yields directly:

\begin{theo} \label{tgc}
Let ${\cal M}$ be a PFW with $M$ complete, and fix $\bar x \in M$.
Then,

\begin{enumerate}
\item[(1)] If $-H(x,u)$ is spatially subquadratic then ${\cal M}$
is geodesically connected.


\item[(2)] If $-H(x,u)$ is at most quadratic with
$$ -H(x,u) \le  R_0(u) d^2(x,\bar x) + R_1(u) d^{p(u)}(x,\bar x) + R_2(u)
$$
$\forall (x,u) \in M \times \R$, $p(u)<2$, then  $z_0 =
(x_0,v_0,u_0)$, $z_1 = (x_1,v_1,u_1) \in {\cal M}$, $u_0\leq u_1$
can be connected by means of a geodesic whenever
$$
R_0[u_0,u_1] (u_1-u_0)^2 < \pi^2,
$$
where
\[
R_0[u_0, u_1] = {\rm Max}\{R_0(u): u \in [u_0, u_1]\}.
\]
\end{enumerate}

\noindent Moreover, in any of previous cases (1), (2):
\begin{enumerate}
\item[$(a)$] If $z_0 < z_1$ there exists a length-maximizing causal geodesic connecting $z_0$ and $z_1$;

\item[$(b)$] If $M$ is not contractible:
\begin{enumerate}
\item[$(i)$] There exist infinitely many spacelike geodesics
connecting $z_0$ and $z_1$,

\item[$(ii)$] The number of timelike geodesics from $z_0$ to $z_v=(x_1,v,u_1)$
goes to infinity when $-v\rightarrow \infty$. 
\end{enumerate}
\end{enumerate}

\end{theo}
It must be emphasized that these results are  optimal because the Riemannian results are optimal too. In fact:
\bit
\item There are explicit counterexamples if any of the hypotheses is dropped.

\item In the case of gravitational plane waves, the conclusions of Theorem \ref{tgc} not only generalize Ehrlich-Emch's ones, but also yield bounds for the appearance of the first astigmatic conjugate pair -a lower bound is the value $u_+$ ($u_+ >u_0$) such that   $R_0[u_0,u_+] (u_+-u_0)^2 = \pi^2$.

\item All the results can be extended naturally to the case $M$ non-complete with convex boundary.
\eit

\subsection{Conjugate points}

From the above approach to geodesic connectedness, it is also
clear that, now, the existence of conjugate points for geodesics
on a PFW is equivalent to the existence of conjugate points for
the action ${\cal J}_\Delta$. More precisely, following
\cite[Section 6]{FScqg}, we can define:

\begin{defi}\label{deff} Fix $\overline{z}_{0}=(x_{0},u_{0})$, $\overline{z}_{1}=(x_{1},u_{1}) \in M \times \R$, and let $x(s)$ be a critical point of ${\cal J}_{\Delta}$ in (\ref{jdelta}) with endpoints $x_{0},x_{1}$ and $\Delta u = u_1-u_0$. We say that  $\overline{z}_{0}$, $\overline{z}_{1}$ are conjugate points along $x(s)$ of multiplicity $\overline{m}$ if the dimension of the nullity of the Hessian of ${\cal J}_{\Delta}$ on $x(s)$ is $\overline{m}$ (if $\overline{m}=0$ we say that $\overline{z}_{0}$, $\overline{z}_{1}$ are not conjugate).
\end{defi}
Then, one obtains the following equivalence between conjugate points for Lorentzian geodesics and conjugate points for Riemannian trajectories of a potential \cite[Proposition 6.2]{FScqg}:
\begin{propo}\label{reeff} The pairs $\overline{z}_{0}=(x_{0},u_{0})$, $\overline{z}_{1}=(x_{1},u_{1})$ are conjugate of multiplicity $\overline{m}$ along $x(s)$, if and only if for any  geodesic  $z: [0,1]\rightarrow {\cal M}$ with $z(s)=(x(s),v(s),\Delta u\cdot s+u_{0})$ the corresponding endpoints $z_{0}=(x_{0},v_{0},u_{0})$, $z_{1}=(x_{1},v_{1},u_{1})$ are conjugate with the same multiplicity $m=\overline{m}$.
\end{propo}
As we commented in Subsection \ref{s5a}, in the particular case of
gravitational plane waves, conjugate pairs are defined for $u_0,
u_1$. For general PFW's, the lack of symmetries of the fronts
makes necessary to take care of the $M$ part. Nevertheless, the
dependence in $v$ is still dropped.

Now, studying the conjugate points for ${\cal J}_\Delta$, one can
obtain easily results as \cite[Proposition 6.4]{FScqg}: {\em if
$H$ is spatially convex (i.e. Hess$_xH[x,s](w,w)\geq 0$, $\forall
w \in TM$) and the sectional curvature of $M$ is non-positive then
no geodesic admits conjugate points. } Of course, the hypotheses
of this result go in the {\em wrong} direction respect to the
energy conditions ($\Delta_xH \leq 0, K_0\geq 0$)), which tend to
yield conjugate points. Nevertheless, this focusing is, in
general, qualitatively different to  the focusing in the plane
wave case and, as we have seen, it does not forbid global
hyperbolicity.

\section*{Appendix: the Riemannian problem of connectedness by the trajectories of a Lagrangian}

In Section \ref{s5} we show that geodesic connectedness of PFW's
depends crucially on the Riemannian variational result Theorem
\ref{trp}. This result is an answer to classical Bolza problem,
which can be stated as:

\begin{quote}
{\bf Bolza problem.} Fixed $x_0 , x_1$ in a Riemannian manifold $M$ and some $T > 0$, determine the existence of critical points for the functional:
$$J_T(x) = {1\over 2} \int_0^T \langle\dot x,\dot x\rangle_{R} ds
 -\int_0^T V(x,s) ds$$
 on the set of  absolutely continuous curves with
$x(0)=x_0,  x(T)=x_1.$
\end{quote}
In our case, $T=1$, $V$ is smooth and at most quadratic, and $M$ is complete.
About this problem, it is well-known that two abstract conditions on $J_T$, namely, boundedness from below and coercitivity, imply the existence of a critical point -in fact a minimum. Even more, by using Ljusternik-Schnirelmann theory one can ensure the existence of a sequence of critical points  such that  $J_T$ diverges.

The following results were known:
\begin{enumerate}
\item If $V(x,s)$ is bounded from above or subquadratic in $x$, then the two abstract conditions hold and $J_T$ attains a minimum.

\item If $V(x,s)$ is at most quadratic, with $V(x,s) \leq \lambda
d^2(x,\bar x)  + \mu d^{p(s)}(x,\bar x) + k(s), p(s)<2, \forall
s\in [0,T]$ then: \bit  \item Clarke and  Ekeland \cite{CE} proved
that, if $T < 1/\sqrt{\lambda}$ then $J_T$  still admits  a
minimum. \item If $T \ge \pi/\sqrt{2\lambda}$ there are  simple
counterexamples to the existence of critical points (harmonic
oscillator).

\eit
\end{enumerate}
Therefore, there was a gap
for the values of $\lambda$,
$$\lambda \in [1/\sqrt{\lambda}, \pi/\sqrt{2\lambda)},$$
which was  covered only in some particular cases (for example, if Hess$_xV \geq 2 \lambda $, then $J_T$ still admits a minimum).
Our results in \cite{CFSjde} (essentially, Theorem \ref{trp}) fill this gap,  by showing that, even in the case $\lambda \in [1/\sqrt{\lambda}, \pi/\sqrt{2\lambda)}$, functional $J_T$ is bounded from below and coercitive and, thus, admits a minimum.

The proof was carried out in three steps:

\bit \item Step 1. The essential term to prove the abstract conditions for $J_T$ is $d^2(x(s),\bar x)$. Then, consider the new functional
$$
F_{T}^\lambda(x) = \frac 1 2\ \int_0^T \langle\dot x,\dot
x\rangle_{R}\ ds \ -\ \lambda\ \int_0^T d^2(x(s),\bar x)\ ds.
$$
 $J_T$ is essentially greater than $F^{\lambda}_{T}$ and, if
$F^{\lambda}_{T}$ is bounded from below and coercitive, then so is
$J_T$ (recall that the expression of $F^{\lambda}_{T}$ contains
$d^2(\cdot,\bar x)$, which is only continuous, but we are not
looking for critical points of this functional).

\item Step 2. Reduction to a problem in one variable.
For each curve $x(s)$ in the domain of $J_T$, one can find a continuous curve $y(s), s\in
[0,T]$, almost everywhere differentiable, such that $y(0)=0$, $y(T)=d(x_0,x_1)$ and:
$$\dot y(s) = |\dot{x}(s)| \hbox{ a.e. in}\; [0,s_{0}],    \quad  \quad
\dot y(s) = -|\dot{x}(s)| \hbox{ a.e. in}\; ]s_0,T],$$ for some
suitable $s_0$. For this curve $y(s)$,
\be \label{ineq}
F_{T}^\lambda(x) \ge \frac 1 2\ \int_0^T |\dot y|^2\ ds
\ -\ \lambda\ \int_0^T |y|^2\ ds.
\ee
And, then,  one has just to prove that the new ({\em
1--dimensional}) functional $G_{T}^\lambda(y)$, equal to the right
hand side of (\ref{ineq}), is coercitive and bounded from below.

\item Step 3. Solution of the 1-variable problem for $G_{T}^\lambda(y)$ by elementary
methods (Fourier series, Wirtinger's inequality). \eit 
The
technique also works for manifolds with boundary \cite{CFSwil}.
Remarkably, the procedure has also been used to
prove the geodesic connectedness of static spacetimes under
critical quadratic hypotheses \cite{BCFS} (see also
\cite{Sa-non}), and other problems.

\section*{Acknowledgments}
J.L.F. has been supported by a MECyD Grant EX-2002-0612. M.S. has
been partially supported by a Spanish MCyT-FEDER Grant
BFM2001-2871-C04-01.




\end{document}